\documentstyle[prl,aps,epsf,multicol,rotate]{revtex}
\title{
Manifold self-localization in a deformable medium
}
\author{
Eugene B. Kolomeisky$^{1}$ and Joseph P. Straley$^{2}$}
\address{
$^{1}$Department of Physics, University of Virginia, Charlottesville, Virginia 22901 \\
$^{2}$Department of Physics and Astronomy, University of Kentucky,Lexington, Kentucky 40506-0055 }
\begin{document}
\maketitle
\begin{abstract}
Directed manifolds (domain walls, interfaces, vortex lines) in a 
deformable medium can exist in a correlated state in which the 
manifold is self-localized by its own strain field.  Depending on 
the temperature, manifold/medium dimensionalities, and the 
strength of the coupling with the medium, the degree of 
localization of the ground state can vary both continuously and 
discontinuously; there can be phase transitions from self-
localized to the free-manifold state.
\end{abstract}
\vspace{-0.75cm}
\begin{multicols}{2}
\narrowtext

The concept of a {\it manifold }is a central paradigm in condensed 
matter physics, covering the cases of domain walls in 
ferroelectric and ferromagnetic materials, flux lines in type-II 
superconductors, interfaces, and many more\cite{CL}.  Manifolds exist 
in symmetry-broken states of various systems, and occur as 
translationally-invariant topological solutions of the 
corresponding field theories.

In this work we show that the {\it deformability }of a real system can 
bring the manifold into a new correlated {\it self-localized }state 
(SL) which forms through the following scenario: a deformable 
continuum responds to the presence of the manifold by a strain 
field decaying away from the manifold; this field interacts with 
the manifold, suppressing its displacements and as a result the 
manifold can be bound to the strain field.  Self-localization 
takes place only if the overall free energy of the system with 
the SL manifold is less than that of the free manifold in the 
strain-free medium.

We start from two examples demonstrating the physics of self-
localization.

{\it 1.  A domain wall in a two dimensional elastic medium}.  The 
position of the wall relative to axes x and t can be represented 
by the function $x = h(t)$.  The interactions of the domain wall
with the elastic medium can then be modeled by a Hamiltonian in
which the wall is represented as an elastic string of stiffness
coefficient m:
\begin{eqnarray} 
H = {\frac12}m\int dt(dh/dt)^{2} - V\int dt [\partial u/\partial x(x = h)] 
\nonumber
\\
 +  {\frac12}\int dxdt [A(\partial u/\partial t)^{2} + B(\partial u/\partial 
x)^{2}]
\end{eqnarray}
The deformations of the medium give rise to the last term; 
for simplicity these have been described by a scalar 
displacement field u, with A and B being elastic constants.  The 
second term provides the coupling between the 
deformation and the wall, in terms of the strain evaluated at the 
wall position x = h.  Since $-\partial u/\partial x$ is proportional to the local
density change, phenomenologically the coupling $-V(\partial u/\partial h)$ is the
lowest order term in the expansion of  the interaction in powers
of the variable part of the medium density.  The coupling
constant V is independent of the elastic properties of the
medium, and can be estimated in terms of the bulk phase
transition temperature $T_{c}$ and lattice spacing {\it a} as $|V|a \cong  T_{c}$.  
In
writing the coupling in the form  $-V(\partial u/\partial h)$ it was assumed that
we are not too close to the bulk critical point so that the range
of the interaction between the wall and an individual atom is of
order {\it a}, which is invisibly small in the long-wavelength theory
we are using and thus does not appear; however the scale {\it a} will
provide a short-distance cutoff implicitly present in Eq. (1).

Placing the wall at the origin $h = 0$, and
minimizing Eq. (1), we find $\partial u/\partial x = (V/B)\delta (x)$ and  $u(x) =
(V/2B) sign (x)$, which means that at $T = 0$  the strain field is localized at
the domain wall (the true range of the delta-function is the
lattice spacing {\it a}).  The magnitude of the maximal strain can be
estimated as $|V|/Ba = |V|a/Ba^{2} \cong  T_{c}/Ba^{2}$; it is small compared to
unity, since $T_{c}$ is much smaller than the binding energy per atom
$Ba^{2}$; this justifies the use of linear elasticity theory.

At a finite temperature T, the wall and the medium degrees of 
freedom each fluctuate strongly - for the uncoupled system $(V =
0)$ the mean-square fluctuations diverge with the system size {\it L} as
$<h^{2}> \cong  TL/m$, and $<u^{2}> \cong  [T/\sqrt{AB}] \ln (L/a)$,  
respectively$^{2}$.  To
understand the system at finite temperature, below
we develop a variational theory valid to zero order in $<u^{2}>/<h^{2}>
($i.e. ignoring medium fluctuations), and having the spirit of
Flory-type theories of polymer physics\cite{DG}.

Assume that the magnitude of the medium deformation is of order 
u$_{0}$ in a region of size L perpendicular to the wall and negligible 
elsewhere.  The corresponding strain field creates a localizing 
potential for the wall [second term of Eq. (1)].  Both the 
localizing potential and the wall itself fluctuate; however, for 
a macroscopic system the medium fluctuations are much weaker than 
those of the free wall, and so for a first approximation we can 
ignore medium fluctuations altogether, and consider the domain 
wall to be fluctuating in a fixed average self-consistent 
potential due to the strain in the medium.  The localizing 
potential confines the domain wall within a region of typical 
size L.  This diminishes the available number of wall 
configurations, thus decreasing the entropy and increasing the 
free energy.  The corresponding confinement free energy\cite{FF} is 
given by $T^{2}/mL^{2}$, so that the total variational free energy  (per
unit length in the $t$ direction) is estimated as
\begin{equation}
F(L,u_{0}) \cong  {T^{2}\over mL^{2}} - {Vu_{0}\over L} + {Bu_{0}^{2}\over 2L}
\end{equation}
where the second term is the free energy gain of localization 
[corresponding to the second term of Eq. (1), while the last term 
is the increase in the elastic energy due to the creation of the 
strained region near the wall [the last term of Eq. (1)].  The 
variational parameters L and $u_{0}$ can be found by minimizing Eq.(2).
The minimum over $u_{0}$ is given  by $u_{0} = V/B$, and then Eq.(2) becomes
\begin{equation}
F(L, u_{0} = V/B) \cong  {T^{2}\over mL^{2}} - {V^{2}\over 2BL} 
\end{equation}
This expression (shown schematically in Figure 1a) has a minimum 
for finite L and $F_{\min} < 0$, which implies that despite  strong
thermal fluctuations an elastic strain {\it always} accompanies the
domain wall; the strain is localized within an equilibrium
localization length given by
\begin{equation}
L_{eq} \cong  T^{2}B/mV^{2}
\end{equation}
We will call the domain wall (plus the accompanying strain field) 
{\it self-localized }because the wall is "attached" to the strain field 
it generates - the probability of finding the wall a distance x 
away from the strain peak is significant for $|x| \le  L_{eq}$,
negligible elsewhere, and goes to zero as $|x| \rightarrow  \infty $.  Note that
self-localization does not imply any breaking of translational
symmetry: both the wall and the strain field can be shifted
simultaneously without energy cost.  Fluctuations of the SL wall
will necessarily involve the degrees of freedom of the medium;
therefore the SL wall possesses a larger stiffness than the same
free wall in an undeformable medium.

The continuum result (4) is valid whenever the 
localization length $L_{eq}$ is much bigger than the lattice spacing 
{\it a}.  Correspondingly we will distinguish between the cases $L_{eq} \gg 
a$ (the {\it macroscopically SL state}) and $L_{eq} \cong a$ ({\it microscopically SL}). 
These meet at the
crossover temperature $T* \cong  |V| \sqrt{ma/B} \ll  T_{c}.$
For $T
\le  T*$ we have $L_{eq} \cong  a$.  Since the domain wall is an anisotropic
object, it is also localized in the $t$ direction over a different
length scale $L_{t}$, which can be estimated from $L_{eq}^{2} \cong  TL_{t}/$m.  At
the crossover temperature T*, we find $L_{t}* \cong  
ma^{2}/T*
\cong  (T_{c}/T*)a \gg  a$.  On length scales smaller than $L_{t}$ the domain wall
is essentially decoupled from the continuum, and so the
fluctuations of the medium at that scale can be estimated as $<u^{2}>
\cong  [T/\sqrt{AB}] \ln (L_{t}/a)$.  Our theory will be valid whenever 
$<u^{2}>$,
evaluated at the smallest allowed scale $L_{t}*$, is much smaller than
$a^{2}$.  This gives rise to the inequality
\begin{equation}
{T*\over a^{2}\sqrt{AB}} \ln (ma/T*) \ll  1
\end{equation}
For most practical purposes $(A \cong  B)$ the condition (5) is
satisfied exceptionally well, because $T* \ll  T_{c} \ll  Ba^{2}$.  Only when
there are special reasons for $A$ to be very small (a quasi-one-
dimensional medium, for example), do we have to start worrying
about the role of the medium fluctuations.  For every case
consider below the neglect of the medium fluctuations can be
justified on similar grounds.

The consequences of manifold self-localization depend on the 
dimensionalities of the manifold and the continuum as 
demonstrated by the next example.

{\it 2.  A flux line (directed polymer) of stiffness m in a three-
dimensional crystal.  } 
We will assume that the strains are localized within a region of 
size $L$ around the vortex.   
The energy cost of the elastic distortion  [the last term in Eq.(2)].  
is now estimated as
$BL^{2}(u_{0}/L)^{2} = Bu_{0}^{2}$ where $B$ is some combination of the elastic
constants of the crystal.  As a result, the total variational
free energy per unit length will have the form
\begin{equation}
F(L, u_{0}) \cong  {T^{2}\over mL^{2}} - {Vu_{0}\over L} + {Bu_{0}^{2}\over 2}
\end{equation}
The minimum value with respect to u$_{0}$ is
\begin{equation}
F(L, u_{0} = V/BL) \cong  {T^{2}\over mL^{2}} - {V^{2}\over 2BL^{2}}
\end{equation}
In contrast with Eq. (3), both terms have the same scaling 
dependence on $L$, and the outcome depends on the relative size of 
the coefficients.  The two terms are of similar size at the 
critical temperature
\begin{equation}
T_{loc} \cong  \sqrt{mV^{2}/B} ;
\end{equation}
for $T > T_{loc}$ the free energy (7) is minimized for the free 
vortex $(L = \infty )$ in the strain-free medium $(u_{0} = 0)$,  while for $T <
T_{loc}$ the free energy is minimized as $L \rightarrow  0$.   Since the 
strain
(which is of order $u_{0}/L \cong  V/BL^{2})$ diverges as $L \rightarrow  0$, this
indicates that the linear elasticity theory used to estimate the
last term of Eq.(6) fails below the localization temperature
(8).  There are several equivalent ways to cure this problem.

First, we can add anharmonic terms to the right-hand side of 
Eq.(6) which will guarantee that the equilibrium L cannot fall 
below some microscopic scale for $T < T_{loc}$.  Alternatively we can
introduce a short-length cutoff {\it a} and continue to use Eqs.(6)
and (7), but restrict the minimization to the region $L \ge  a$.
Neither of these procedures provides a quantitative picture for $T
< T_{loc}$ but for the  purposes of illustration we note that using
the cutoff (we adopt it hereafter) predicts that for $T \le  T_{loc}$ the
free energy density is given by
\begin{equation}
F \cong  (T^{2} - T_{loc}^{2} )/ma^{2}
\end{equation}
In contrast to the case of a domain wall in two dimensions, a 
flux line in a three-dimensional crystal can exist either in the 
high-temperature free state or in the low-temperature 
microscopically SL phase.  The theory developed here is valid 
only for temperatures well below the superconductive temperature 
$T_{c}$; since $T_{c} \ll  Ba^{3}$, one can deduce from (8) that $T_{loc} 
\ll  T_{c}$.

The nature of these results -- having the same $1/L^{2}$
dependence of both the "localizing" and "delocalizing" terms of
the free energy (7), plus the fact that the expression for
the phase transition temperature (8) is cutoff-independent --
shows that three dimensions is the {\it lower critical
dimensionality}\cite{CL} of the self-localization problem.

The critical behavior (9) predicted by our variational theory 
is only an estimate, and a more detailed theory is needed to 
reveal the essential singularities that are expected to occur at 
the critical point in a marginal case.  The long-ranged nature of 
bulk elasticity is clearly important here: the equilibrium medium 
displacement away from the vortex depends on the variational 
localization length L as $u_{0} \cong  V/BL [$see  (7)], implying that
the radial component of the displacement field depends inversely
$(u \cong  V/Br)$ on the distance r from the vortex.

{\it 3}.  Now let us consider the general case of {\it a D-dimensional 
manifold of stiffness m placed in a d-dimensional continuum.}

For D $<$ 2 the variational free-energy density generalizing 
Eqs.(2) and (6) has the form
\begin{equation}
F(L,u_{0}) \cong  T({T\over mL^{2}})^{D/(2-D)} - {Vu_{0}\over L} + {Bu_{0}^{2} \over 2L^{D+2-d}}
\end{equation}
where the L-dependence of the first term\cite{FF} demonstrates the 
general tendency that thermal fluctuations become less important 
with increasing manifold dimensionality D.  The case  D = 2 is 
the {\it upper critical dimensionality}\cite{FF} for thermal fluctuations 
and will be treated separately.  For $D > 2$ thermal fluctuations
are unimportant, and the free energy density will be given by
just the last two terms of (10).  Eq.(10) includes several
important special cases, such as a vortex line or directed
polymer $(D = 1)$, and a domain wall $(D = d - 1)$, correspondingly
placed in a d-dimensional elastic continuum characterized by an
elastic constant B.

Eq. (10) is minimized (with respect to $u_{0})$ for $u_{min} = V/BL^{d-1-D}$,
which immediately implies that only for the case of a domain wall
$(D = d - 1)$ is the induced strain field well-localized; in all
other cases it is long-ranged.  The corresponding minimum value
of (10) is
\begin{equation}
F(L, u_{0} = u_{min}) \cong  T({T \over mL^{2}})^{D/(2-D)}- {V^{2}\over 2BL^{d-D}}
\end{equation}
Here we have several cases to consider:

For D $>$ 2 (or for any D and $T = 0)$ the first term of (11)
will have to be dropped, and what is left is minimal for $L \rightarrow  0$.
We must assume there is a short-range cutoff {\it a}, and conclude that
the manifold will be in a microscopically SL state with $L_{eq} \cong  a$.

For $2D/(2 - D) > d - D$ and $D < 2$ the right-hand side of (11)
(shown schematically in Figure 1a) has a minimum at
\begin{equation}
L_{eq} \cong  [(BT/V^{2})^{2-D} (T/m)^{D}]^{1/[2D-(d-D)(2-D)]}
\end{equation}
which generalizes Eq.(4).  At the minimum one has $F_{\min} < 0$,
which implies that the manifold is in a macroscopically SL state
as long as $L_{eq} \gg  a$.  Should the localization length $L_{eq}$ decrease
(due to some change in the system parameters), the system
continuously crosses over into a microscopically SL state
characterized by $L_{eq} \cong  a$.

The case $2D/(2 - D) = d - D$ and $D < 2$, similar to the previously
analyzed problem of a flux line, is marginal.  Here the manifold
can exist either in a high-temperature free state or in a low-
temperature microscopically SL phase separated by a continuous
phase transition.

For $2D/(2 - D) < d - D$ and $D < 2$ the right-hand side of (11)
(shown schematically in Figure 1b) has a maximum at a scale $L_{\max}
[$actually given by (12)], approaches zero from above as $L \rightarrow 
\infty $, vanishes at a scale $L_{0} < L_{\max} [$up to a scale factor $L_{0}$ is
still given by the right-hand side of (12)], and goes to
minus infinity as $L \rightarrow  0$.  Here the outcome depends on the
relationship between the cutoff {\it a}, and the scales $L_{0}$ and $L_{\max}$.

For $a > L_{\max}$ (the high-temperature limit) the free-energy density 
(with the cutoff) has a single minimum at $L = \infty $: the manifold is
in a free state.

For $L_{0} < a < L_{\max}$ (intermediate temperatures) the free energy 
density has a global minimum at $L = \infty $, and a local minimum at $L \cong 
a$.  The ground state is still a free manifold, but now a
microscopically SL state can be metastable because of the barrier
separating it from the free state.

For $a < L_{0} $ (a low-temperature limit) the free energy density has 
a global minimum at $L \cong  a$, and a local minimum at $L = \infty $.   The
ground state is a microscopically SL manifold but a free manifold
can be metastable due to the free energy barrier separating
it from the SL state.

The condition $a = L_{0}$ determines a first-order phase transition
point between the microscopically SL phase and a free manifold.

{\it 4.  }For the marginal case D = 2 we will consider only the 
practically most important case of {\it a domain wall in a three-
dimensional crystal.  }The free energy per unit area will now have 
the form
\begin{equation}
F(L, u_{0} = V/B) \cong  (T/a^{2}) e^{- mL^{2}/T} - V^{2}/BL
\end{equation}
where the first term is the free energy of confinement\cite{FF} which 
for  D = 2 replaces the first terms of Eqs.(10-11).

To simplify the analysis it is convenient to introduce the 
dimensionless coupling constant $\gamma  = (V^{2}a^{2}/B)(m/T^{3})^{1/2}$.  For D = 2 
the parameter V is estimated as $|V|a^{2} \cong  T_{c} \cong  ma^{2}$, and the
corresponding coupling constant is $\gamma \cong
(T_{c}/Ba^{3})(T_{c}/T)^{3/2}$, thus implying that for $T < T_{c}$ the parameter 
$\gamma $
can vary from a very small value of order $T_{c}/Ba^{3} \ll  1 ($for a
temperature close to $T_{c})$ to infinity (the zero-temperature
limit).

In the low-temperature limit $(\gamma  \gg  1)$ the function (13) is
monotonic increasing which means that the ground state is a
microscopically SL domain wall.  In the high temperature limit $(\gamma 
\ll  1)$ the right-hand side of (13) (shown schematically in
Figure 1c) has a maximum at some $L_{\max}$, and a minimum at some $L_{\min}
> L_{\max}$.  The transition between the two behaviors occurs at $\gamma  \cong  1$
which corresponds to the temperature $T_{1} \cong  T_{c}(T_{c}/Ba^{3})^{2/3} \ll  
T_{c}$ and
$L_{1} = L_{\max} = L_{\min} \cong  (T_{1}/m)^{1/2}$.  Since the scale $L_{1} \cong  
(T_{1}/m)^{1/2} \cong 
a(T_{1}/T_{c})^{1/2}$ is much smaller than the cutoff {\it a} we conclude that
the domain wall will be in a microscopically SL state $L_{eq} \cong  a$
both for $T < T_{1}$ and in some temperature range above $T_{1}$ until the
cutoff {\it a} coincides for the first time with $L_{\min} (L_{\min}$ increases
with the temperature).  This will happen at some $\gamma  \ll  1$, and in
this limit $L_{\min}$ can be estimated as $L_{\min} \cong  
(T/m)^{1/2}\ln^{1/2}(1/\gamma )$.
The condition $L_{\min} = a$ determines a crossover temperature T* as a
solution to the equation $\ln^{1/2}(1/\gamma ) \cong  a(m/T)^{1/2}$.  Remembering
that $\gamma  = (V^{2}a^{2}/B)(m/T^{3})^{1/}{ } ^{2}\cong  
(T_{c}/Ba^{3})(T_{c}/T)^{3/2}$, we can solve the
equation with logarithmic accuracy, $\ln(Ba^{3}/T_{c}) \gg  1$, to find $T* \cong 
T_{c}/\ln(Ba^{3}/T_{c})$.
\begin{figure}[thb]
\epsfxsize1.5in
\epsfbox{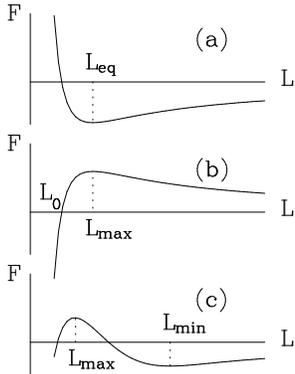}
\caption{
Sketch of the free energy F of the manifold as
a function of the variational localization length L for the
various regimes of manifold and medium dimensionalities:
(a) $2D/(2-D) > d-D,$ $D < 2;$ (b) $2D/(2-D) < d-D,$
$D < 2;$ (c) a domain wall in a three-dimensional crystal
(D=2,d=3) in the high-temperature limit.
}
\end{figure}

The physical significance of the crossover temperature T* is that 
for $T < T*$ the free energy density has a single minimum at $L_{eq} \cong 
a$, and the domain wall is in a microscopically SL state.  At the
same time for $T > T*$ the free energy density has a minimum at

\begin{equation}
L_{eq}^{2} \cong  (T/m) \ln (B^{2}T^{3}/mV^{4}a^{4})
\end{equation}
The ground state is still self-localized but its localization 
length (14) is bigger than {\it a }and grows with temperature.  
The bulk phase transition temperature $T_{c}$ sets a natural upper
bound to $L_{eq}$, which can be estimated as $L_{eq} \le  a \ln^{1/2}(Ba^{3}/T_{c})$.
In reality the dependence (14) is only reliable for $T* < T
< T_{c}$ excluding the temperature range of bulk critical
fluctuations.

Because thermal fluctuations are only marginally relevant for a 
domain wall in a three-dimensional crystal, experimental 
verification of the law (14) (for example, by measuring the 
strain profile induced by the wall) may not be easy unless 
$\ln(Ba^{3}/T_{c}) \gg  1$, and the bulk critical region is narrow.
 
It is known experimentally and well-understood theoretically\cite{LK}
how topological defects and other inhomogeneities deform the
embedding medium at zero temperature.  In our language this corresponds to
the microscopically SL manifold.  As we have shown, thermal fluctuations can
change this picture drastically leading to the possibility of
macroscopically SL states or even strain-free states.  The fact that the
manifold is in a SL state should be especially important when the
accompanied strain field is long-ranged (this excludes domain walls), and
when there are many manifolds present.  We speculate that the SL transition
predicted for a single flux line will manifest itself in a structural
transition of a dilute system of vortices.
The present theory implies that there is a similar phenomenon for any
$D = 1$ manifold in a deformable medium in three dimensions.

The problems analyzed above involve directed manifolds placed in 
a deformable medium.  For the special case of line manifolds $(D =
1)$ they are connected to quantum-mechanical phenomena at T = 0
through the Feynman formulation of quantum mechanics\cite{Kogut}.
The relationship goes through the correspondence temperature $\rightarrow $
Planck's constant, line configuration $\rightarrow $ particle world line,
position along the line $\rightarrow $ imaginary time, and line stiffness 
$\rightarrow $
particle mass.  Then line self-localization phenomena are
formally related to the polaron effect in quantum physics - an
electron placed in an elastic continuum might get self-trapped by
its own strain field\cite{Rashba}.  More precisely, the Hamiltonian
Eq. (1) can be viewed as an imaginary-time Action for the polaron
problem in one dimension, while the expression for the
localization length (4) can be related to the size of a large
polaron in the adiabatic limit\cite{Newb}.
This implies that the SL manifold has a well-localized
probability
distribution function which depends only on transverse coordinates.
This is 
qualitatively different from the free manifold, which has
a Gaussian probability  
distribution function that depends on all coordinates.

\end{multicols}
\end{document}